\documentclass[8.5pt,twoside,twocolumn]{article}
\usepackage[super,sort&compress,comma]{natbib} 
\usepackage{mhchem}
\usepackage{times,mathptmx}
\usepackage{sectsty}
\usepackage{balance} 

\usepackage{graphicx} 
\usepackage{lastpage}
\usepackage[format=plain,justification=raggedright,singlelinecheck=false,font=small,labelfont=bf,labelsep=space]{caption} 
\usepackage{fancyhdr}
\begin{document}

\thispagestyle{plain}
\fancypagestyle{plain}{
\fancyhead[L]{\includegraphics[height=8pt]{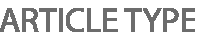}}
\fancyhead[C]{\hspace{-1cm}\includegraphics[height=20pt]{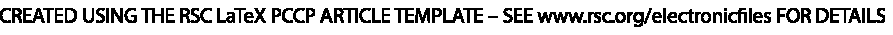}}
\fancyhead[R]{\includegraphics[height=10pt]{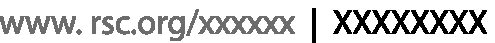}\vspace{-0.2cm}}
\renewcommand{\headrulewidth}{1pt}}
\renewcommand{\thefootnote}{\fnsymbol{footnote}}
\renewcommand\footnoterule{\vspace*{1pt}%
\hrule width 3.4in height 0.4pt \vspace*{5pt}} 
\setcounter{secnumdepth}{5}

\makeatletter 
\def\subsubsection{\@startsection{subsubsection}{3}{10pt}{-1.25ex plus -1ex minus -.1ex}{0ex plus 0ex}{\normalsize\bf}} 
\def\paragraph{\@startsection{paragraph}{4}{10pt}{-1.25ex plus -1ex minus -.1ex}{0ex plus 0ex}{\normalsize\textit}} 
\renewcommand\@biblabel[1]{#1}            
\renewcommand\@makefntext[1]%
{\noindent\makebox[0pt][r]{\@thefnmark\,}#1}
\makeatother 
\renewcommand{\figurename}{\small{Fig.}~}
\sectionfont{\large}
\subsectionfont{\normalsize} 

\fancyfoot{}
\fancyfoot[LO,RE]{\vspace{-7pt}\includegraphics[height=9pt]{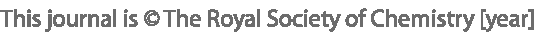}}
\fancyfoot[CO]{\vspace{-7.2pt}\hspace{12.2cm}\includegraphics{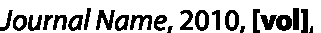}}
\fancyfoot[CE]{\vspace{-7.5pt}\hspace{-13.5cm}\includegraphics{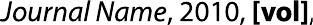}}
\fancyfoot[RO]{\footnotesize{\sffamily{1--\pageref{LastPage} ~\textbar  \hspace{2pt}\thepage}}}
\fancyfoot[LE]{\footnotesize{\sffamily{\thepage~\textbar\hspace{3.45cm} 1--\pageref{LastPage}}}}
\fancyhead{}
\renewcommand{\headrulewidth}{1pt} 
\renewcommand{\footrulewidth}{1pt}
\setlength{\arrayrulewidth}{1pt}
\setlength{\columnsep}{6.5mm}
\setlength\bibsep{1pt}

\twocolumn[
  \begin{@twocolumnfalse}
    \noindent\LARGE{\textbf{Velocity Force Curves, Laning, and Jamming for
        Oppositely Driven Disk Systems }}
\vspace{0.6cm}

\noindent\large{\textbf{C. Reichhardt\textit{$^{a}$} and
C.J.O. Reichhardt$^{\ast}$\textit{$^{a}$}
}}\vspace{0.5cm}

\noindent\textit{\small{\textbf{Received Xth XXXXXXXXXX 20XX, Accepted Xth XXXXXXXXX 20XX\newline
First published on the web Xth XXXXXXXXXX 200X}}}

\noindent \textbf{\small{DOI: 10.1039/b000000x}}
\vspace{0.6cm}

\noindent \normalsize{
Using simulations we examine a two-dimensional disk system in which two disk species are driven in opposite directions.  We measure the average velocity of one of the species versus the applied driving force and identify four phases as function of drive and disk density: a jammed state, a completely phase separated state, a continuously mixing phase, and a laning phase.   The transitions between these phases are correlated with jumps in the velocity-force curves that are similar to the behavior observed at dynamical phase transitions in driven particle systems with quenched disorder such as vortices in type-II superconductors.  In some cases the transitions between phases are associated with negative differential mobility in which the average absolute velocity of either species  decreases with increasing drive.  We also consider the situation where the drive is applied to only one species as well as systems in which both species are driven in the same direction with different drive amplitudes.  Finally, we discuss how the transitions we observe could be related to absorbing phase transitions where a system in a phase separated or laning regime organizes to a state in which contacts between the disks no longer occur and dynamical fluctuations are lost.       
}
\vspace{0.5cm}
  \end{@twocolumnfalse}
]

\section{Introduction}

\footnotetext{\textit{$^(a)$~Theoretical Division,
    Los Alamos National Laboratory, Los Alamos, New Mexico 87545, USA.
    Fax: 1 505 606 0917; Tel: 1 505 665 1134; E-mail: cjrx@lanl.gov}}

A wide variety of systems can be modeled as a collection of interacting particles 
that, when driven over a quenched substrate, exhibit depinning and dynamical
transitions as a function of increasing driving force \cite{1}. Such systems include 
vortices in type-II superconductors \cite{2,3,4}, electron crystals \cite{5},
driven colloidal systems \cite{6,7,8,N,9}, and sliding friction \cite{10,11}. 
At low drives these systems are in a pinned
state where the velocity is zero, while above a critical driving force, 
the particles become depinned and slide. Within the
moving states there can be different dynamical modes of motion such as 
a plastic phase with strong fluctuations in 
the particle positions and velocities\cite{1,2,4,6,7}. 
At higher drives the system can 
organize to a dynamically ordered state such as a moving crystal \cite{2,3,12}  
or moving smectic \cite{4,12,13,14,15}. For particles driven 
over a periodic substrate, additional types of dynamic phases can appear such 
as soliton motion or one-dimensional (1D) to
two-dimensional (2D) transitions,
along with ordered and disordered flow phases \cite{1,8,N,9,10,11,16,17,18}.  
The transitions between these different dynamical states are associated
with cusps, jumps, or dips in 
the velocity force curves,
as well as with global changes in the ordering of the particle configurations
or the amount of dynamical fluctuations. 
In all these systems the pinning arises from quenched disorder that
is fixed in space; however, there can also be cases
where the pinning is not fixed but can move in response to the 
driven particles.
For example, if a number of particles that are not coupled to the
external drive can block the motion of 
particles that are coupled to the external drive,
the driven particles can move the blocking particles and over time rearrange 
them to create a new landscape or pattern \cite{19,20}. 
A previously studied system that closely resembles this case is
two species of interacting particles driven in opposite directions that
exhibit a variety of 
dynamical behavior, including a transition to a laning state \cite{21,22,23,24,25,26,27}
where the particles separate into
quasi one-dimensional chains of the same species, as well as regimes in which
the particles mix and undergo disordered flow.
Such phases have been observed
in experiments on
colloids moving in opposite directions \cite{28} and dusty plasma systems \cite{29}.
This type of system can also exhibit
pattern forming states \cite{24,25,30,31} and jammed or clogged states \cite{24,32}.
Certain active matter systems
have similar behavior, such as pedestrian flows 
that can be mimicked by  particles moving in opposite directions \cite{33}. 

In this work we examine a two-dimensional system of disks 
in the absence of thermal fluctuations.
Half the disks are driven 
in one direction and the other half in the opposite direction,
and we measure the net velocity
of one disk species along with the amount of six-fold ordering
as a function of the driving force and disk density. 
Previous studies of laning transitions have generally focused on systems  
of colloids or particles with Yukawa interactions \cite{21,23}.  In our case
the system is close to the hard disk limit
and the density $\phi$ is defined as the area covered by the 
disks.  For $\phi > 0.9$ the 
disks form a triangular solid or jammed state \cite{34}.
For $\phi \geq 0.55$ we find that the disks can organize into four possible 
dynamic phases: a jammed phase (I) where all the disks
are in contact forming a triangular solid with zero net velocity;
a fully phase separated state (II) where the disks organize into two bands 
with crystalline order  
moving in opposite directions and disk-disk collisions do not occur;
a strongly fluctuating disordered phase (III)
where disk collisions are continuously occurring and the system has liquid like features;
and a laning state (IV)
where the disks form a series of lanes, disk-disk collisions are absent,
velocity fluctuations drop to zero, and the system has smectic properties.
The transitions between these phases correlate 
with changes in the net velocity of either species as well
as with changes in the disk ordering and the nature of
the dynamic fluctuations.
The mobility in the phase separated and laning states
is high since the disks can move freely past one another without collisions,
while the transition to the
disordered phase is accompanied by a drop in the net velocity, leading to 
a region of negative differential mobility similar to that found
in transitions from laminar flow to turbulent flow as a function of
increasing drive in certain systems with quenched disorder \cite{16,17,18}. 
For $\phi < 0.55$ the system 
always organizes into a laning state where all disk collisions are lost. 
We also examine the situation where only one species is driven and show that the
same four phases can arise in the high density limit.
Here the jammed state consists of a
drifting solid phase
where the non-driven disks lock to the driven disks, while
the phase separated and laning states are composed of assemblies
of driven disks moving past stationary regions of 
non-driven disks.   
These phases can even occur when both species are driven in the same direction with
different couplings to the external drive. 
We conjecture that some of these transitions fall into the class of 
absorbing phase transitions \cite{35} 
when the system reaches a state where the disk collisions and fluctuations are
completely lost, similar to the
recently observed irreversible-reversible
transitions in periodically sheared disk systems \cite{36}.

\section{Simulation}
We model a 2D system of size $L \times L$ with $L=36.0$
with periodic boundary conditions  
in the $x$ and $y$ directions 
containing $N_d$ disks of radius $R_d=0.5$. 
The disk-disk interaction is modeled as a repulsive
harmonic spring.
The overdamped equation of motion for a disk $i$ is  
\begin{equation}
\eta \frac{d {\bf R}_{i}}{dt} = {\bf F}^{i}_{dd}  + {\bf F}_{D}^i .
\end{equation}
The disk-disk interaction force is 
${\bf F}_{dd} = \sum_{i\neq j}^{N_d}k(2R_{d} - |{\bf r}_{ij}|)\Theta(2R_{d} - |{\bf r}_{ij}|) {\hat {\bf r}_{ij}}$,
where ${\bf r}_{ij} = {\bf R}_{i} - {\bf R}_{j}$,
$\hat {\bf r}_{ij}  = {\bf r}_{ij}/|{\bf r}_{ij}|$, and $\Theta$ is the Heaviside
step function.
The spring constant $k = 60$, and we
find negligible changes in the dynamics for larger values of $k$.
Each disk is subjected to an applied driving force
${\bf F}_D^i= A^iF_D{\bf \hat{x}}$ where $N_1$ disks
are driven in the positive $x$ direction with
$A^i = C$, with $C=1.0$ unless otherwise noted.
The remaining $N_2=N_d-N_1$ disks
are driven in the negative $x$ direction with $A^i = -1.0$.  
After applying the drive,
we wait for the system to settle into 
a steady state.
This transient waiting time is a strong function of disk density, and under
some conditions can be as large as
$1 \times 10^8$ simulation time steps.
After the system
reaches a steady state we measure 
the average disk velocity for each species and normalize
it by $N_{1(2)}$
to obtain the average velocity per disk  
$\langle V_{1(2)}\rangle = N_{1(2)}^{-1}\sum^{N_{1(2)}}_{i=1}{\bf v}_{i}\cdot {\hat{\bf x} }$,
where ${\bf v}_{i}$ is the
instantaneous velocity of disk $i$.  Since we use overdamped dynamics with a 
damping constant of $\eta=1.0$,
in the free flow limit the disks move at a velocity of
$\langle V_{1}\rangle=F_{D}$ and $\langle V_{2}\rangle=-F_{D}$.
The density $\phi$ is defined 
as the area coverage of the disks,
$\phi = N_{d}\pi R^2_{d}/L^2$, and in the absence of driving
the system forms a uniform crystalline solid at $\phi = 0.9$.

\section{Velocity Force Curves and Dynamic Phases}

\begin{figure}
 \centering
\includegraphics[width=0.5\textwidth]{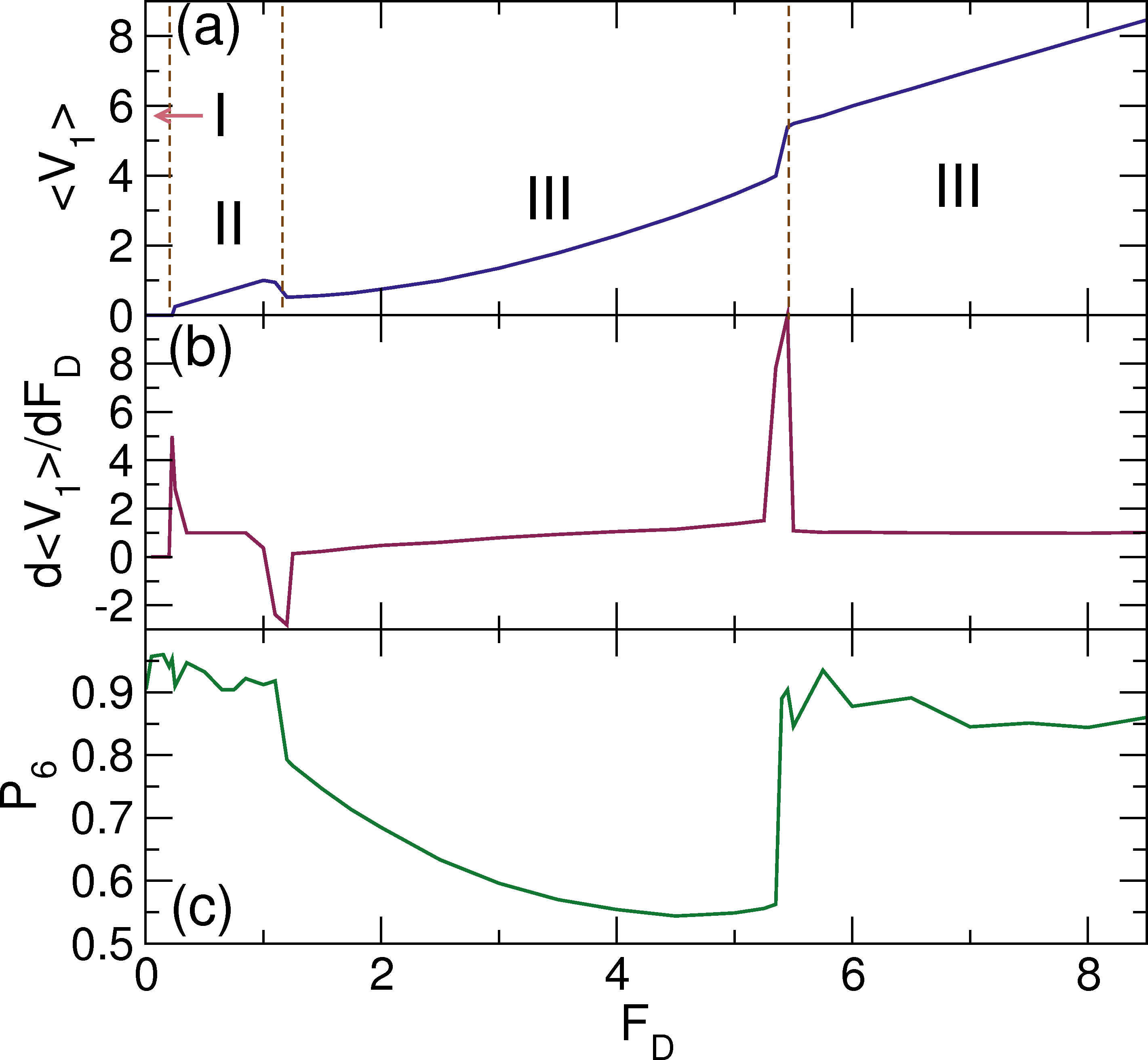}
\caption{(a) The average velocity per disk $\langle V_1\rangle$, measuring only
  the disks driven in the $+x$ direction,  vs $F_{D}$
  for a system with oppositely driven disks and $N_1=N_2=0.5N_d$ at
  $\phi = 0.848$. 
  (b) $d\langle V_1\rangle/dF_{D}$ vs $F_{D}$ for the same system.
(c) The corresponding fraction of sixfold coordinated disks $P_{6}$ vs $F_D$.
We find four phases: I (jammed), II (phase separated), III (disordered flow),
and IV (laning).
The transitions between the states appear as jumps or dips
in the various measures.
}
\label{fig:1}
\end{figure}

\begin{figure}
 \centering
\includegraphics[width=0.45\textwidth]{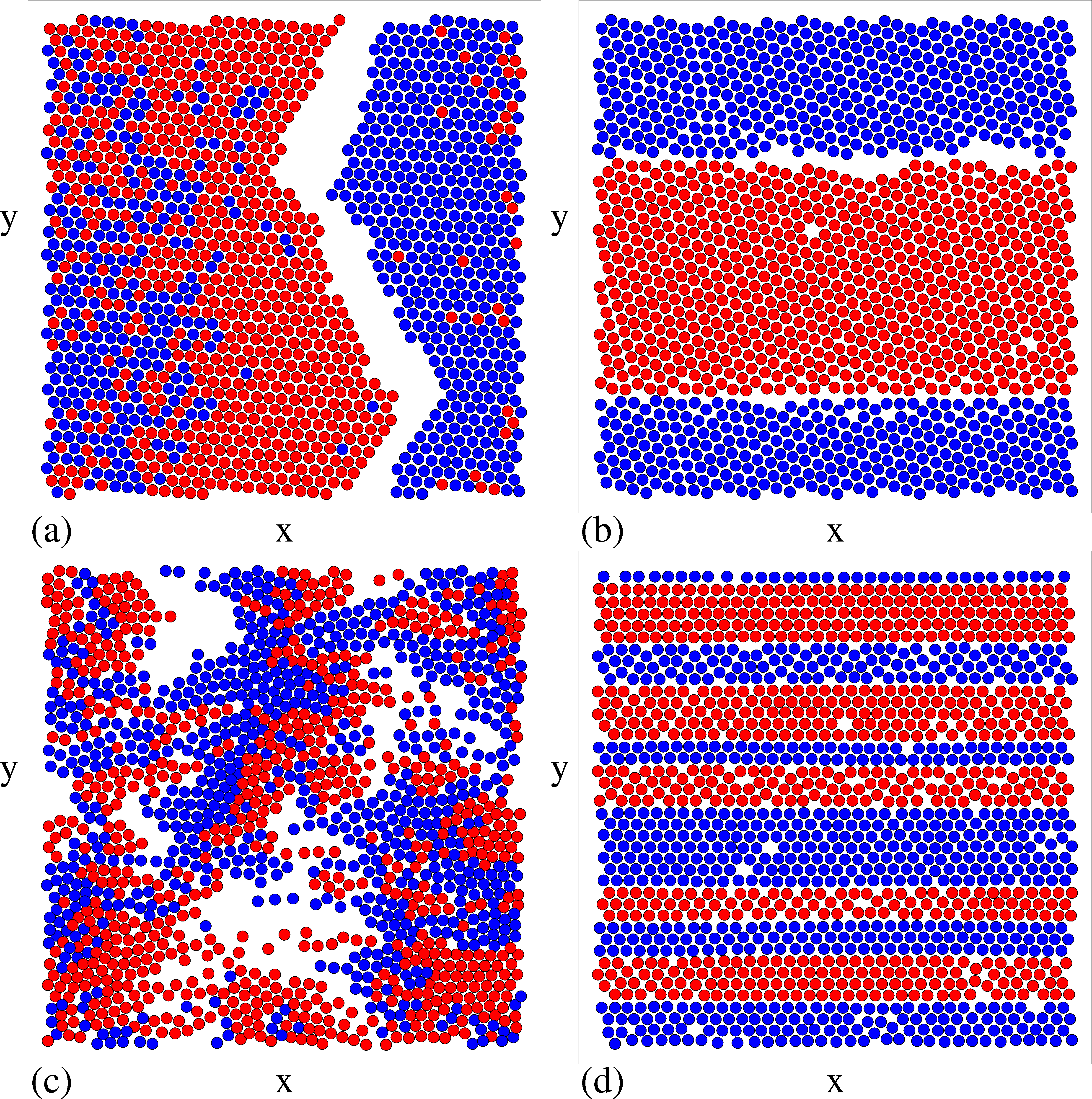}
\caption{The disk configurations from the system in Fig.~\ref{fig:1}.
The blue disks (species 1) are driven in the
$+x$ direction and the red disks (species 2) are driven
in the $-x$ direction. (a) The jammed phase I at $F_{D} = 0.15$.
(b) The phase separated state II at $F_{D} = 0.75$. (c) The disordered phase III
at $F_{D} = 3.0$. (d) The laning phase IV at $F_{D} = 6.5$.
}
\label{fig:2}
\end{figure}

We first consider samples in which $N_1=N_2=0.5N_d$.
In Fig.~\ref{fig:1}(a) we plot $\langle V_1\rangle$ versus $F_{D}$ 
for the disks driven in the $+x$ direction 
for a system with $\phi = 0.848$, and in Fig.~\ref{fig:1}(b) we show
$d\langle V_1\rangle/dF_D$ versus $F_D$.
We also measure the fraction $P_6$ of sixfold coordinated disks for all disks,
$P_6=\sum_{i}^{N_d}\delta(z_i-6)$ where the coordination number $z_i$ of disk $i$ is
obtained using a Voronoi construction, and plot $P_6$ versus $F_D$ in
Fig.~\ref{fig:1}(c).
The corresponding $\langle V_2\rangle$ versus $F_{D}$
curve for disks driven in the $-x$ direction 
looks exactly the same as the curve in Fig.~\ref{fig:1}(a) but is negative.

We identify four distinct dynamic phases 
based on transitions in the velocity-force curve.
The jammed phase (I)
appears for $0 < F_{D} < 0.3$ and has
no disk motion, $\langle V_1\rangle=0$, and
strong sixfold disk ordering, $P_{6} \approx 0.95$. 
In phase I, the
disks form a dense cluster with triangular order, as illustrated in 
the disk configuration image in Fig.~\ref{fig:2}(a) 
for $F_{D} = 0.15$.
Within the jammed phase the local disk density $\phi_{\rm loc}$
is close to $\phi_{\rm loc}=0.9$, and since this is lower than the total disk density,
there is a small region containing no particles ($\phi_{\rm loc}=0$). 
We note that in phase I $P_{6}<1$
since the disks on the edge of the jammed cluster do not have six neighbors. 

At $F_{D} = 0.3$, a jump in $\langle V_1\rangle$ and
a peak in $d\langle V_1\rangle/dF_{D}$ indicate
the transition from phase I to phase II,
which is similar to the 
peak in $d\langle V\rangle/dF_{D}$ observed in systems with
quenched disorder at the pinned to 
sliding transition \cite{1}. 
Phase II, the phase separated state, extends over the range $0.3 < F_{D} < 1.15$,
and in this phase $\langle V_1\rangle$ increases linearly with
increasing $F_{D}$ and 
$d\langle V_1\rangle/F_{D} = 1.0$,
indicating that the particles are in a free flow regime. 
Additionally, $P_{6} \approx 0.92$ and
both species exhibit triangular ordering as shown in Fig.~\ref{fig:2}(b) at $F_{D} = 0.75$.

At $F_{D} = 1.15$ we find a transition from phase II to phase III
accompanied by a drop in $\langle V_1\rangle$
which produces a negative spike in $d\langle V_1\rangle/dF_{D}$. 
This is an example of negative differential mobility where the
disk velocity decreases with increasing drive.
The II-III transition is also associated with a drop in $P_{6}$ when
the system enters the disordered flow phase.
In systems with 
quenched disorder, negative differential mobility has also been
reported at transitions from ordered
to disordered or turbulent flow phases \cite{10,16,17,18}.
Phase III appears for $1.15 < F_{D} < 5.45$,
and is characterized by strongly fluctuating structures of the type
shown in Fig.~\ref{fig:2}(c) for $F_{D} = 3.0$.
The two particle species
are strongly mixed and the sixfold ordering is lost.
The disks continuously undergo collisions and show strong
velocity deviations in the $y$ direction, transverse to the drive.
Within phase III, $P_{6}$ gradually decreases to $P_6\approx 0.55$ near 
$F_{D} = 5.45$.

The transition from phase III to phase IV  appears
as an upward jump in both $\langle V_1\rangle$
and $P_{6}$ along with a positive peak in
$d\langle V_1\rangle/dF_{D}$.
As shown in Fig.~\ref{fig:2}(d) at $F_D=6.5$, 
in phase IV
the disks form multiple oppositely moving lanes.
There is considerable triangular ordering of the disks
within each lane which produces the
increase in $P_{6}$ at the III-IV transition.
For higher values of $F_{D}$, the system
maintains phase IV flow
and $d\langle V_1\rangle/dF_{D} = 1.0$,
indicating that the disks are in a free flow regime.     

\begin{figure}
 \centering
\includegraphics[width=0.5\textwidth]{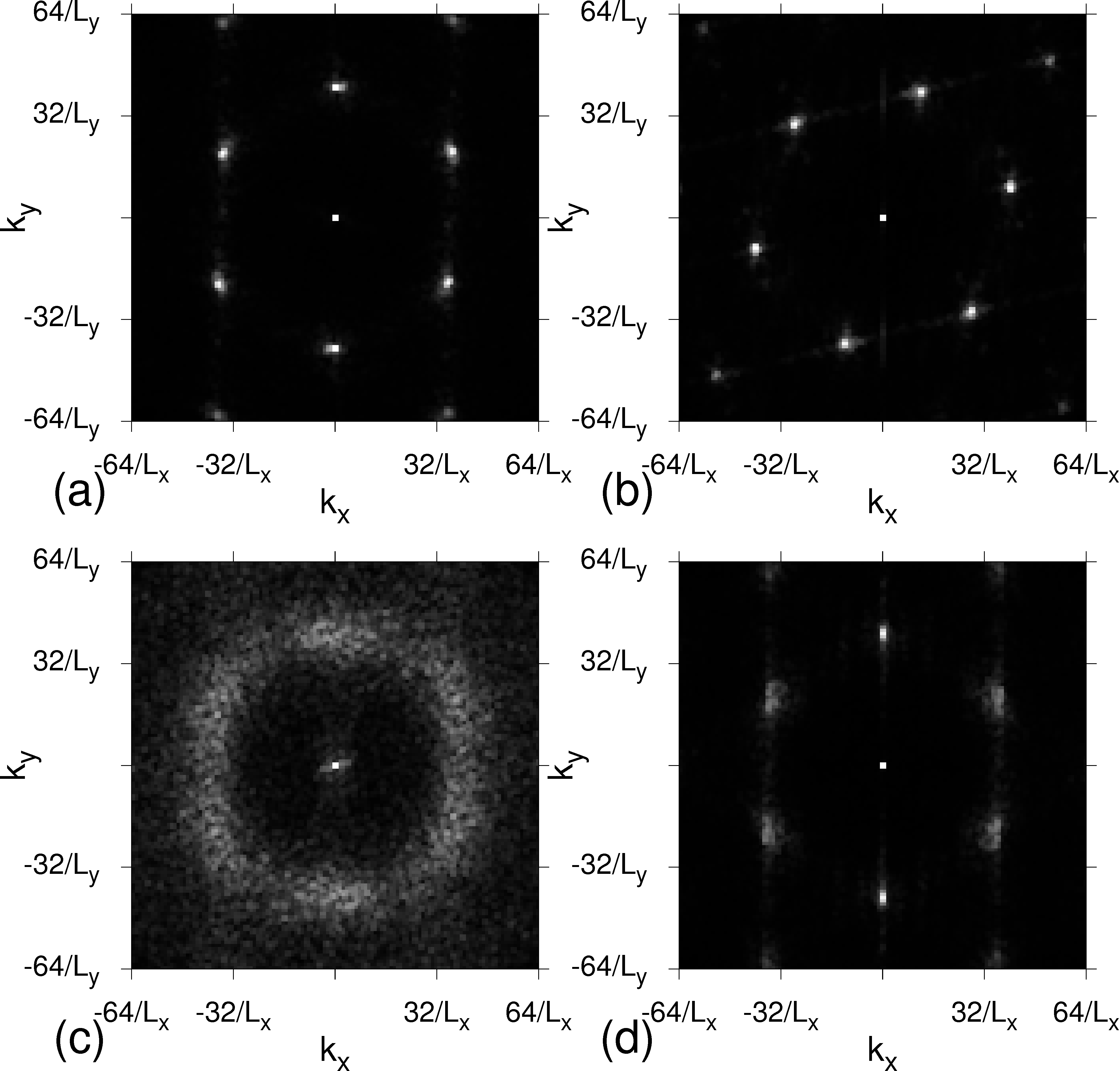}
\caption{ The structure factor $S({\bf k})$ for the four phases in Fig.~\ref{fig:2}.
  (a) The jammed phase I at $F_D=0.15$ has triangular ordering.
  (b) The phase separated state II at $F_D=0.75$ has triangular ordering.
  (c) The disordered flow phase III at $F_D=3.0$
  shows a ring shape indicating liquid ordering.
  (d) The laning phase IV at $F_D=6.5$
  has a smectic character with weak triangular ordering.
}
\label{fig:3}
\end{figure}

We can characterize the structure of the disks in the different phases
using the  
structure factor
$S({\bf k})=N_d^{-1}|\sum_i^{N_d}\exp(-i{\bf k} \cdot {\bf r}_i)|^2 $.
In Fig.~\ref{fig:3}(a) we plot $S({\bf k})$ for
the system in Fig.~\ref{fig:2}(a) in phase I at $F_{D} = 0.75$, 
where we find six peaks indicative of triangular ordering.
In phase II, Fig.~\ref{fig:3}(b)
shows a sixfold pattern of peaks with
a small amount of smearing of the peaks.
The disordered flow phase III in
Fig.~\ref{fig:3}(c)
has a ring pattern indicative
of liquid ordering,
while in Fig.~\ref{fig:3}(d), the laning phase IV
has two strong peaks at $k_{x} = 0.0$ and four weaker side peaks consistent with
a moving smectic structure containing some local triangular ordering.  
The changes in $P_{6}$ and $S({\bf k})$ from a liquid structure to a moving smectic
signature
as a function of increasing $F_{D}$ are similar to the transitions observed
for particles
moving over quenched disorder 
such as vortices in type-II superconductors \cite{2,3,14,15}.

\begin{figure}
 \centering
\includegraphics[width=0.5\textwidth]{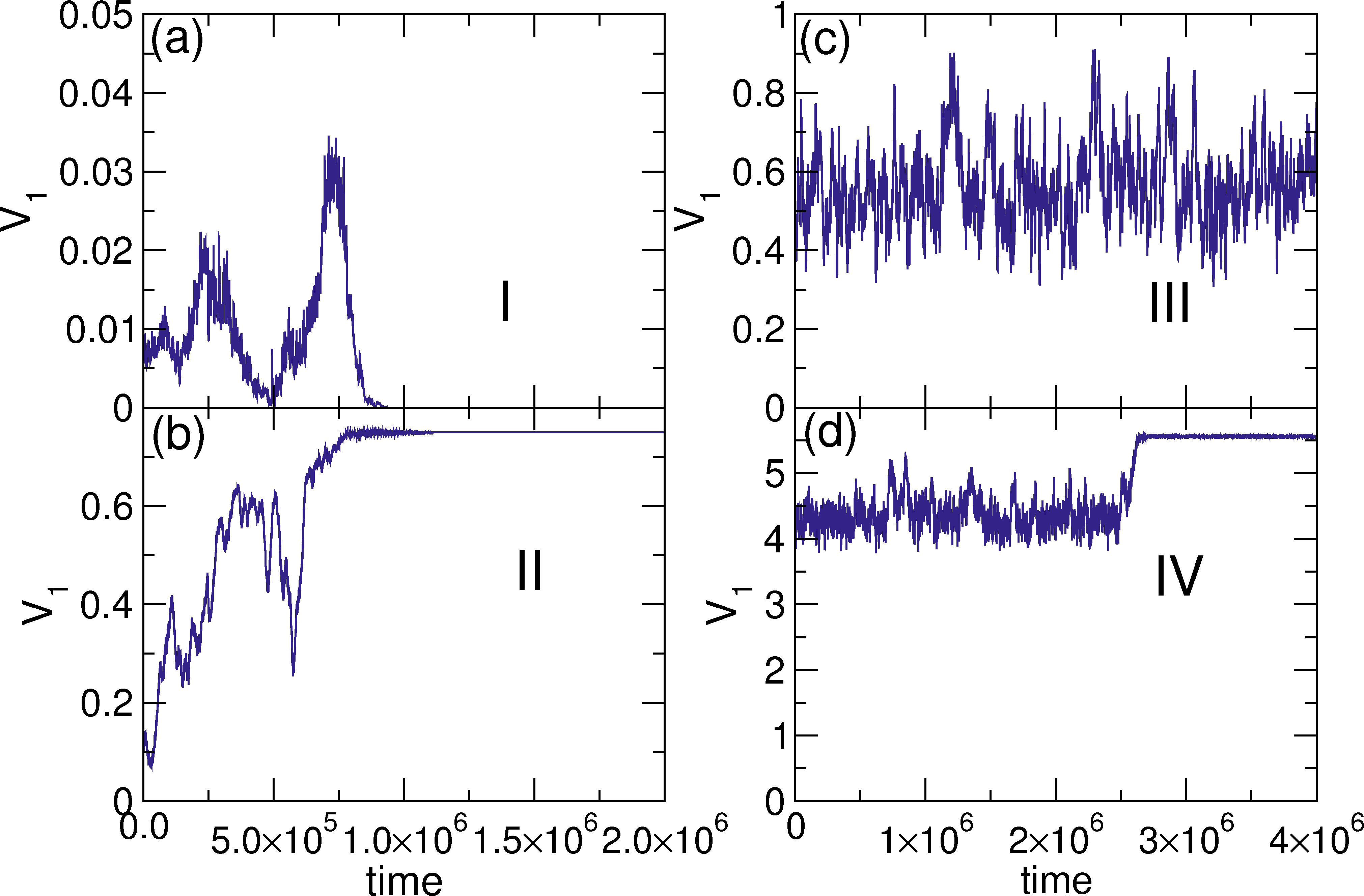}
\caption{The instantaneous velocity $V_1$ per disk vs time
  for species 1 for the system in Fig.~\ref{fig:1} at $\phi = 0.848$.
  (a) Phase I at $F_{D} = 0.15$, where $V_1$ goes to zero. 
  (b) Phase II at $F_{D} = 0.75$,
  where $V_1$ saturates to a fluctuation-free flowing state with $V_1 = 0.75$. 
  (c) Phase III at $F_{D} = 1.5$,
  where the system remains in a strongly fluctuating state with
  $\langle V_1\rangle = 0.56$.
  (d) Phase IV at $F_{D} = 5.57$,
  where the system is initially in a fluctuating state and organizes
  at later times into a fluctuation-free flowing state
  with $V_1 = 5.57$.    
 }
\label{fig:4}
\end{figure}

We can get insight into the dynamic fluctuations of the
different phases by examining the time series of the 
instantaneous velocity $V_1(t)$ of species 1 disks in the different phases. 
In Fig.~\ref{fig:4}(a) we plot $V_1$ versus
time in phase I at $F_{D} = 0.15$.
Initially $V_1$ is in 
a fluctuating transient state indicating that the disks are moving,
but at later times 
the system organizes into a jammed state with
$V_1 = 0$. 
In phase III at $F_D=0.75$, Fig.~\ref{fig:4}(b) shows
that there are initially strong fluctuations in $V_1$ but that at later times
the system settles into a fluctuation-free
flowing state with $V_1 = 0.75$.
This corresponds to the formation of the 
phase separated state, where $V_2=-0.75$.
The absence of fluctuations in 
$V_1$ and the fact that $V_1=F_D$
indicate that the disks are in a completely free flow state 
and that disk-disk collisions
no longer occur.
In Fig.~\ref{fig:4}(c) for
phase III at  $F_{D} = 1.5$,
$V_1$ strongly fluctuates between
$V_1=0.4$ and $V_1=0.9$, and the
average velocity $\langle V_1\rangle = 0.56$ is almost three 
times smaller than the free flow value of $\langle V_1\rangle=1.5$.
In this phase, the velocity continues to fluctuate out to the longest simulation times
we consider, and the strong fluctuations indicate that there are continuous disk-disk
collisions that impede the flow of the disks in both directions.
Figure~\ref{fig:4}(d) shows phase IV flow at $F_{D} = 5.57$,
where the system is initially in a fluctuating flow
phase similar to phase III, but organizes at later times 
into a non-fluctuating free flow state
where all disk-disk collisions are lost.

\begin{figure}
\includegraphics[width=0.5\textwidth]{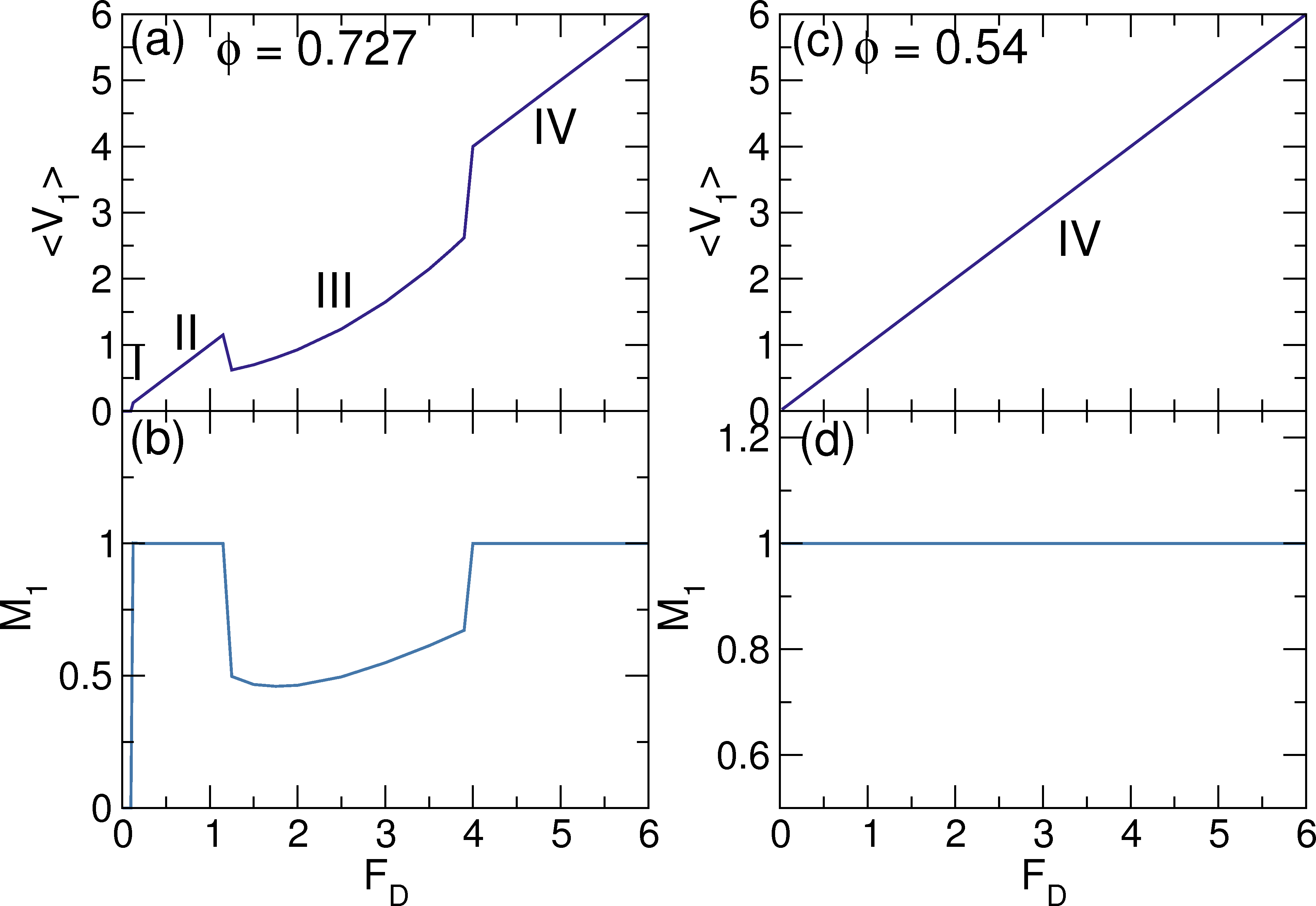}
\caption{(a) $\langle V_1\rangle$ vs $F_{D}$ for a system with
  $N_1=N_2=0.5N_d$
  at $\phi = 0.727$. (b) The corresponding mobility 
  $M_1=\langle V_1\rangle/F_{D}$ vs $F_{D}$
  shows that the I-II and III-IV transitions are shifted to lower values of $F_D$
  compared to the $\phi=0.848$ system.
  (c) $\langle V_1\rangle$ and (d) the mobility $M_1$ vs $F_D$
  for a system with $\phi=0.54$, where the disks are always in phase IV,
  the velocity-force curve is linear, and $M_1=1$.
 }
\label{fig:5}
\end{figure}

We investigate the evolution of the phases as a function of $\phi$ and
find that for decreasing $\phi$ the III-IV transition
drops to lower values of $F_{D}$.
In Fig.~\ref{fig:5}(a) we plot $\langle V_1\rangle$
versus $F_{D}$ for the system in Fig.~\ref{fig:1} 
at $\phi = 0.727$,
while Fig.~\ref{fig:5}(b) shows the corresponding mobility
$M_1=\langle V_1\rangle/F_{D}$ versus $F_{D}$. 
The I-II transition has dropped to $F_D=0.1$
and the mobility in phase I is $M_1=0$.
The II-III transition occurs at roughly the same value
of $F_{D}$ as in the $\phi = 0.848$ system, but 
the III-IV transition drops to $F_{D} = 4.0$.
When the system is in free flow in phases II and IV,
$M_1=1$,
but the mobility is substantially reduced in the disordered flow phase III.
In Fig.~\ref{fig:5}(c,d) we plot
$\langle V_1\rangle$ and the mobility
$M_1$ versus $F_{D}$ for a sample with $\phi = 0.54$, 
where the disks always organize into phase IV flow,
the velocity-force curve is linear,
and $M_1=1$.

\begin{figure}
\includegraphics[width=0.5\textwidth]{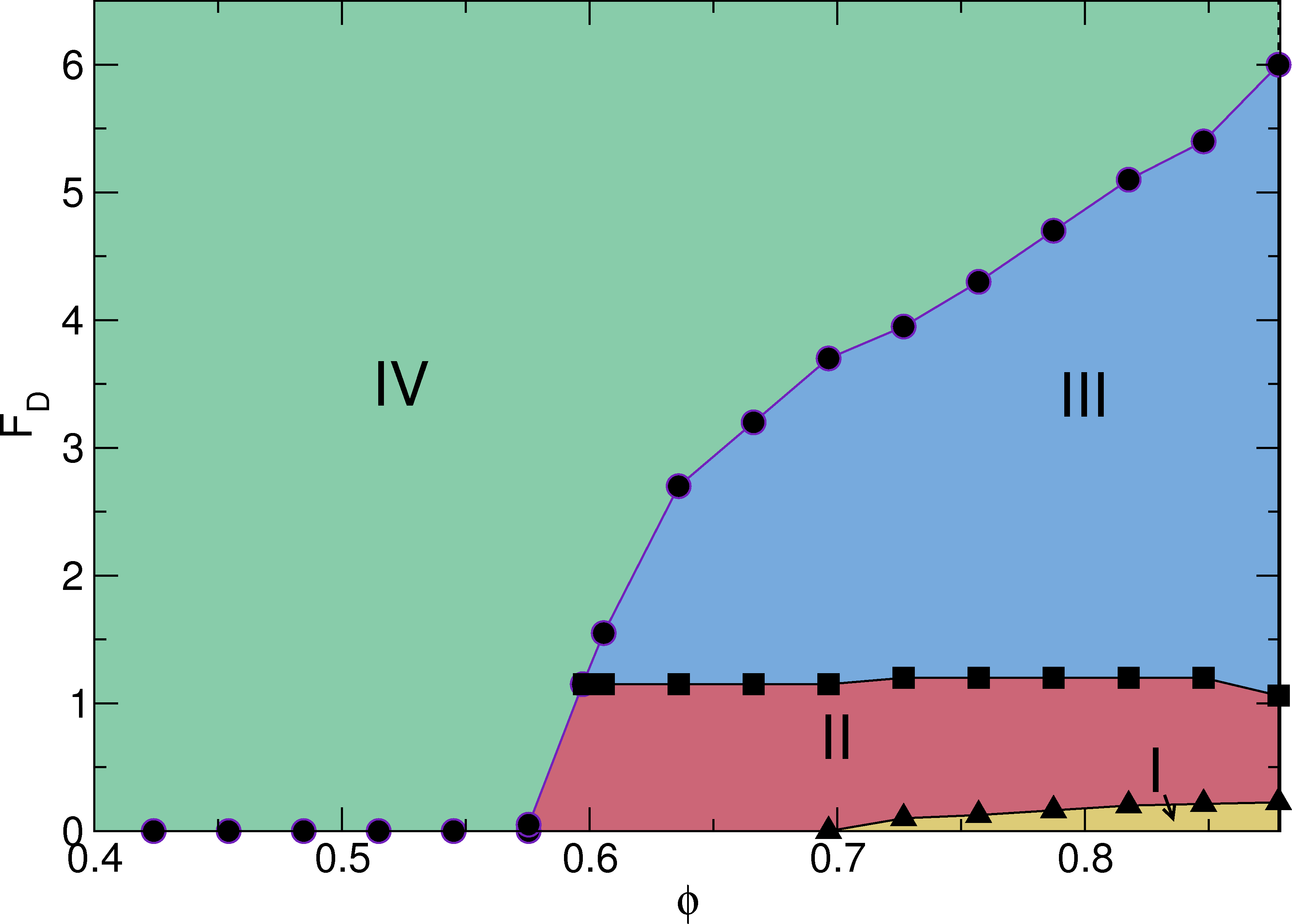}
\caption{Dynamic phase diagram
  as a function of $F_{D}$ vs $\phi$.
  I: jammed phase; II: phase separated state; III: disordered flow phase;
  IV: laning phase.
  For $\phi < 0.55$ the system
is always in phase IV, while phase III grows in extent with increasing $\phi$ 
for $\phi > 0.55$.  
 }
\label{fig:6}
\end{figure}

By conducting a series of simulations we map the evolution of the different phases
as a function of 
$F_{D}$ vs $\phi$, as shown in
the dynamic phase diagram in Fig.~\ref{fig:6}.
For $\phi > 0.55$,
phases II, III, and IV all occur,
while for $\phi < 0.55$ only phase IV is present.
For $\phi > 0.55$, the extent of phase III grows 
with increasing $\phi$,
and we only observe region I for $\phi > 0.7$. 
The II-III transition occurs at roughly the same value of $F_D$ as
$\phi$ varies.
The minimum value of $\phi = 0.55$ at which
phases II and III first appear may
be related to a contact percolation transition.
In compressional simulations
of 2D monodisperse frictionless disks,
Shen {\it et al.} \cite{37} found a contact percolation transition
at $\phi_{p} = 0.549$ and argued that this transition is connected to the onset of
a non-trivial mechanical response or stress in the system.
Several of the phase transitions we observe have similarities to 
the irreversible-reversible transition observed
in periodically sheared disk suspensions, where the system 
can transition from an
irreversible fluctuating state where collisions occur to a reversible
state where collisions are lost \cite{36,38}. 
In our system, in phases II and IV the collisions are lost, while phase III 
would correspond to an irreversible state. 
Additionally, the jammed phase can also be
viewed as an absorbed state since all fluctuations are lost, even though the
particles are now all in contact.  

\section{Varied Species Ratios and Driving Force Ratios}

\begin{figure}
\includegraphics[width=0.5\textwidth]{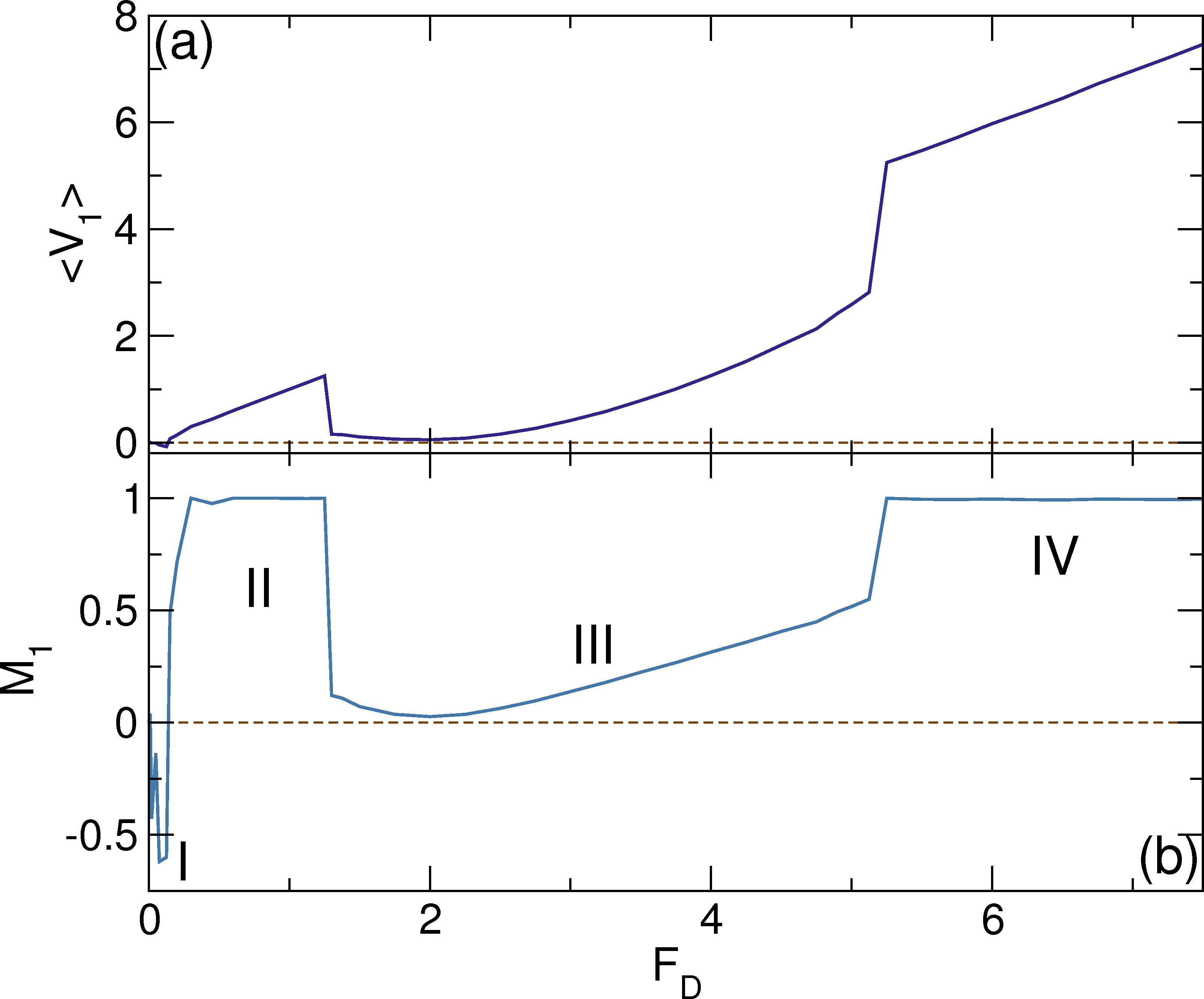}
\caption{ (a) $\langle V_1\rangle$ vs $F_{D}$ for a sample with $\phi=0.848$
  in which $N_1=0.2N_d$ and $N_2=0.8N_d$. 
  (b) The corresponding mobility $M_1$ vs $F_D$.
  Here $M_1=1.0$ in phases II and IV, but
  there is a portion of phase III for which the mobility is nearly zero. 
  Phase I is no longer a jammed phase with
  $\langle V_1\rangle = 0$.   Instead, a rigid cluster
  forms that drifts in the $-x$ direction,
 producing a negative mobility.    
}
\label{fig:7}
\end{figure}

We have also considered different ratios of $N_1$ to $N_2$
and find that the same general phases appear.
In Fig.~\ref{fig:7}(a) we plot $\langle V_1\rangle$
versus $F_{D}$ for a system with $\phi=0.848$, $N_1=0.2N_d$, and $N_2=0.8N_d$, 
while Fig.~\ref{fig:7}(b) shows the corresponding mobility 
$M_1$ versus $F_{D}$.
Here the same four phase arises and the overall shape 
of the curves is similar to that of the $N_1=N_2=0.5N_d$ system
shown in Fig.~\ref{fig:1};  however, in phase III the 
average mobility drops nearly to zero since
the additional collisions suppress the flow of the disks in  
the $+x$ direction.
In phases II and IV, $M_1=1$, indicating that the system 
can still organize into a collisionless
state where the disks undergo free flow motion.
A notable difference is that for phase I in the $N_1=0.2N_d$ sample,
the mobility $M_1$ is negative.
The jammed phase I for the $N_1=0.5N_d$ sample has $\langle V_1\rangle = 0$,
but in the $N_1=0.2N_d$ sample,
the jammed state consists of 
a rigid solid that translates in the 
$-x$ direction since $N_2>N_1$.
In general, for any ratio other than $N_1/N_2=1$,
the jammed phase I has a net drift in the
direction of the majority species.
As $N_1/N_2$ is decreased further,
in phases II and IV we always find $M_1=1$,
while for phases I and III the mobility decreases or becomes more negative.

\begin{figure}
\includegraphics[width=0.5\textwidth]{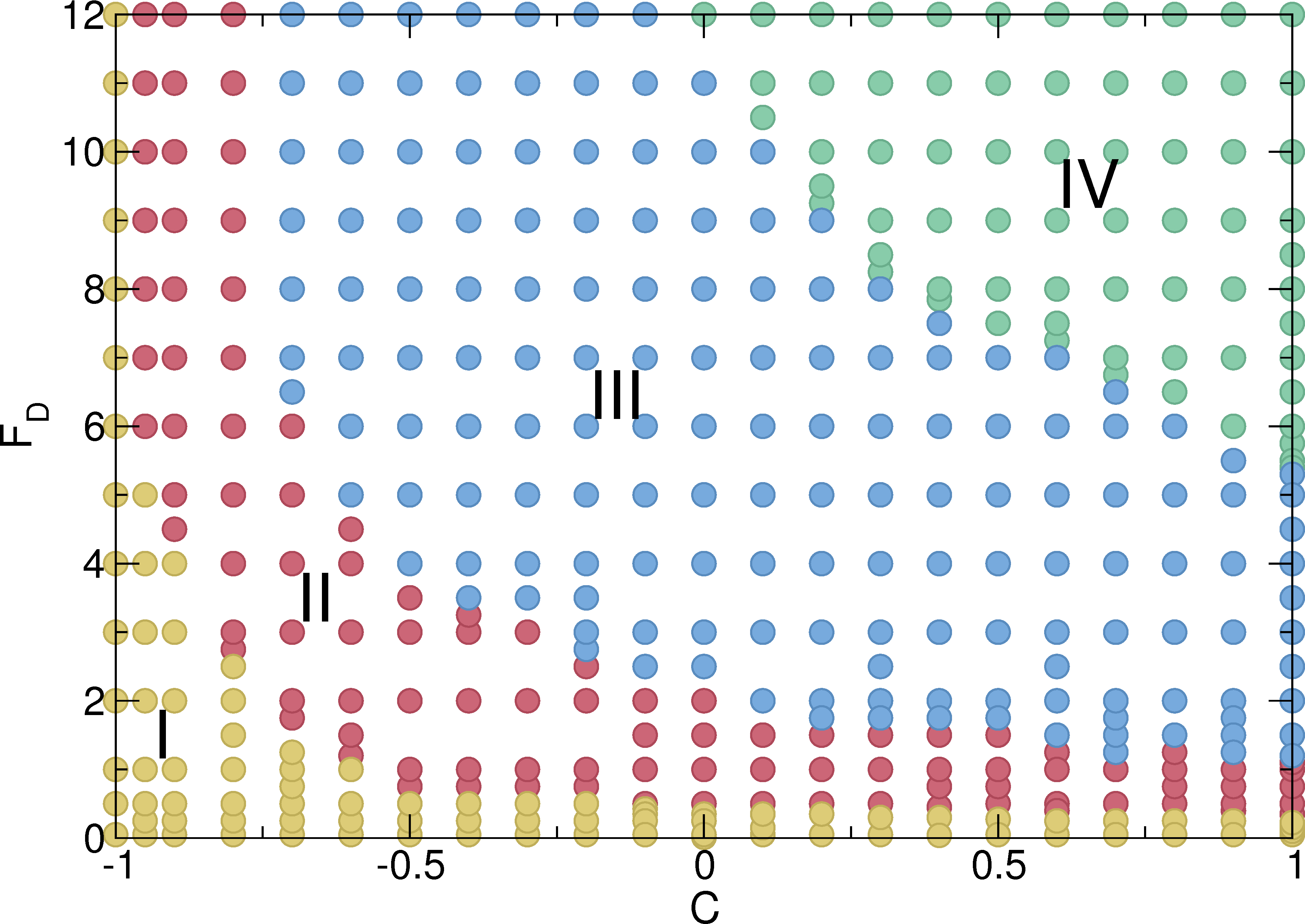}
\caption{Dynamic phase diagram
  as a function of $F_{D}$ vs $C$,
  where $C$ is the coefficient controlling the amplitude of the driving force
  for species 1 disks, in a sample with $N_1=N_2=0.5N_d$ and $\phi=0.848$.
  The value $C = 1.0$ corresponds to the
  case shown in Fig.~\ref{fig:1} for exactly oppositely driven particles.
  Yellow circles: jammed phase I; red circles: phase separated state II;
  blue circles: disordered flow phase III; and green circles: laning phase IV.
At $C= 0$ where only species 2 couples to the drive, all four phases can still occur.
   }
\label{fig:8}
\end{figure}

We can also change the ratio of the relative driving force on the two different species.
We consider a system with $N_1=N_2=0.5N_d$ at $\phi=0.848$ and vary the
value of $C$ controlling the amplitude of the driving force for species 1, while
keeping the drive on species 2 fixed at ${\bf F}_D^i=-F_d{\bf \hat x}$.
Here $C$ is chosen in the range
$ -1.0 \leq C \leq +1.0$.
The exactly oppositely driven disks in Fig.~\ref{fig:1} correspond to the case
$C=+1.0$.
For $C = 0$, half of the disks do not couple to the external drive while the other half are
driven in the negative $x$ direction,
and for $-1.0 \leq C < 0$,
both species are driven in the $-x$ direction 
with different forces.
At $C = -1.0$, all the particles are driven in the $-x$ direction with the
same force.
In Fig.~\ref{fig:8} we show a dynamic phase diagram
as a function of $F_{D}$ versus $C$.
All of the phase transitions shift to higher values of $F_D$ as $C$ decreases.
It is interesting to note that for $C=0$, where only species 2 is driven, all four
dynamic phases still occur.

\begin{figure}
\includegraphics[width=0.5\textwidth]{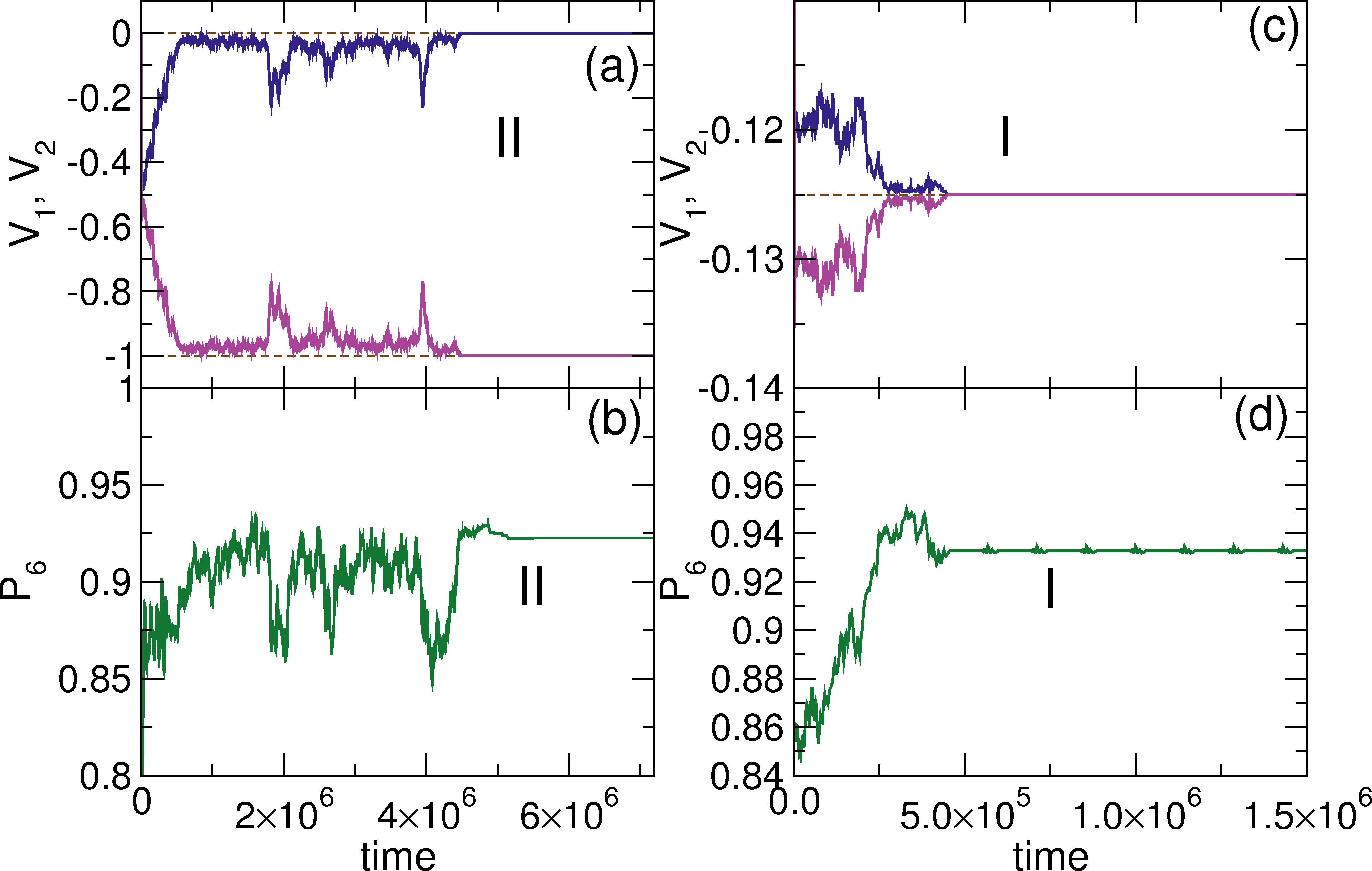}
\caption{A sample from Fig.~\ref{fig:8} with $C=0$ and $F_D=1.0$.
  (a) The instantaneous velocity per disk $V_{1}$ vs time in simulation time steps
  (upper blue curve) for the non-driven disks and
the corresponding
  $V_{2}$ vs time for the disks driven in the $-x$ direction.
  Here the system organizes into
  phase II flow  with $V_{1} = 0$ and $V_{2} = -F_D=-1.0$.
  (b) $P_{6}$ vs time for all the disks in the same system
  showing the transition into phase II.
  (c) $V_1$ and $V_2$ vs time for the same system at $F_{D} = 0.25$ and $C = 0$.
  The disks organize into phase I where both
  species become locked together and move at $V_1=V_2 = -0.125 = F_{D}/2$.
  (d) The corresponding $P_{6}$ vs time curve shows that
  the system organizes into a mostly triangular state at
  the transition to phase I.   
  }
\label{fig:9}
\end{figure}

\begin{figure}
\includegraphics[width=0.5\textwidth]{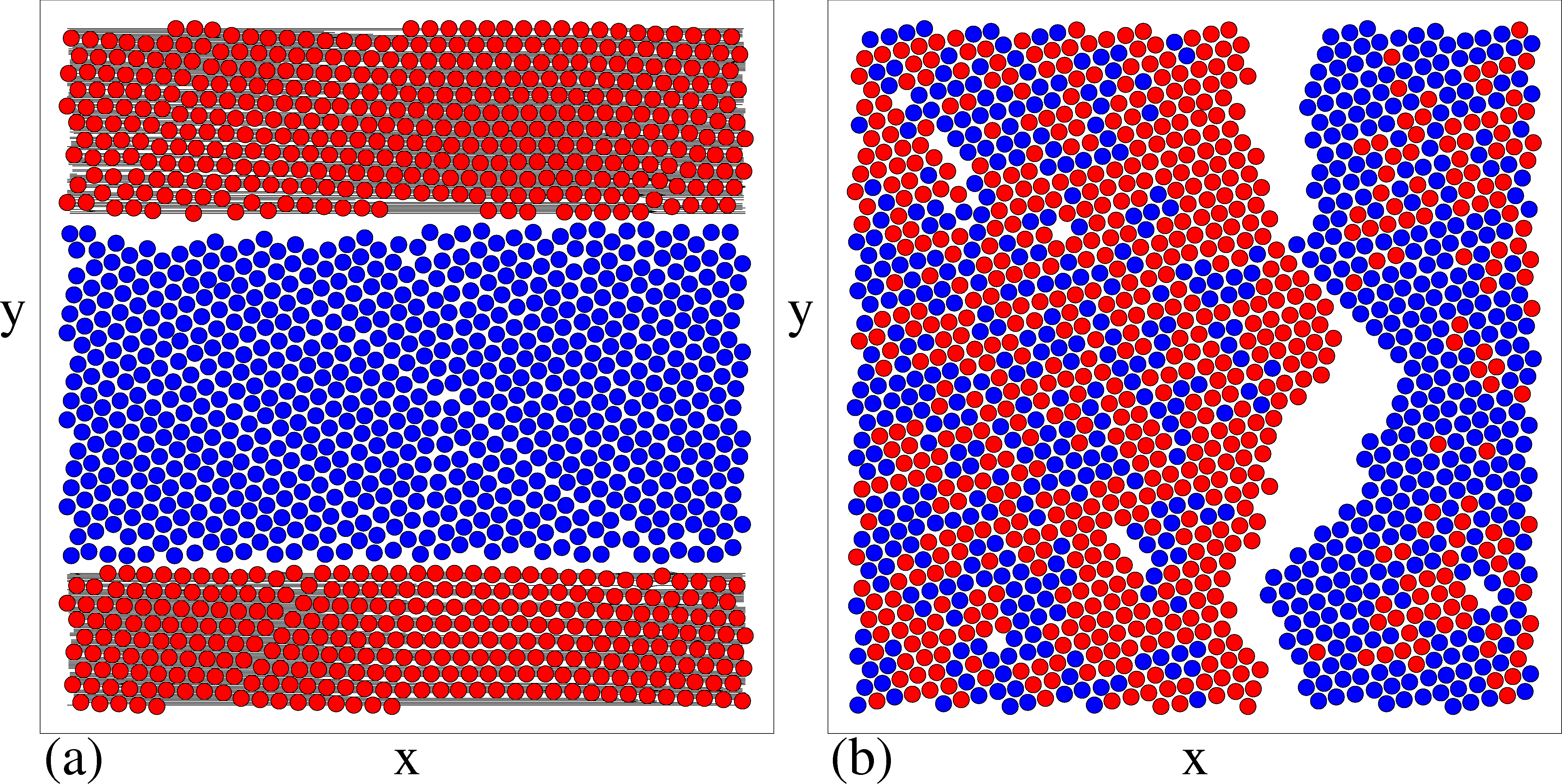}
\caption{ The disk configurations for the system in Fig.~\ref{fig:9}
  with $C=0$.  The
  blue disks (species 1) are not driven and the red disks (species 2) are driven
  in the $-x$ direction.
  (a) The phase separated state II for the system in 
  Fig.~\ref{fig:9}(a,b) at $F_{D} = 1.0$.
  The black lines are the disk trajectories 
  indicating that the system has phase separated into moving and
  non-moving spatial regions.    
  (b) The jammed state I at $F_{D} = 0.25$ from Fig.~\ref{fig:9}(c,d)
  where the entire system is moving in the negative $x$-direction.  
}
\label{fig:10}
\end{figure}

In Fig.~\ref{fig:9}(a) we plot
the instantaneous disk velocities
$V_1$ and $V_2$ versus time for $C = 0$ and $F_D=1.0$.
Here, species 1 is not driven.
The system organizes into phase II as indicated by the transition to
$V_{1}=0$ and $V_{2}=-1.0$.
The velocities rapidly approach the phase II values at short times,
but significant velocity fluctuations persist and appear as
jumps in both $V_{1}$ and $V_{2}$ in the transient fluctuating state. 
In Fig.~\ref{fig:9}(b), the corresponding $P_{6}$ versus $F_{D}$ 
for all the disks shows that the large 
fluctuations in $V_{1,2}$ correlate with drops in $P_{6}$,
and that when the system fully settles into
phase II flow, $P_{6}$ saturates to $P_6=0.925$, indicating mostly triangular ordering.
In Fig.~\ref{fig:10}(a) we show the disk configurations
and trajectories in the phase II flow for the $C=0$ system in Fig.~\ref{fig:9}(a). 
The disks phase separate into 
an immobile triangular packing surrounded by a  moving triangular lattice.
At high drives for $C=0$, a laning phase IV appears where
half of the lanes are immobile and the other half  
are moving.

The jammed phase I at $C=0$ 
consists of a moving solid drifting in the $-x$-direction,
as illustrated 
in Fig.~\ref{fig:9}(c) where we plot
$V_{1}$ and $V_{2}$ versus time at $F_{D} = 0.25$.
Here the velocities of both disk species
converge   
to $V_1=V_2 = -0.125 = -F_{D}/2$
in phase I since the non-driven disks reduce the velocity 
of the driven disks to half of its free-flow value.
Figure~\ref{fig:9}(d) shows the corresponding
$P_{6}$ versus time,  where at the transition into phase I,
$P_{6}=0.93$ indicating the formation of a triangular drifting solid. 
In Fig.~\ref{fig:10}(b) we show the disk configurations in phase I
for the system in Fig.~\ref{fig:9}(c),
where the driven disks 
drag the non-driven disks.

Phases I and II persist 
for  $-1.0 < C < 0.0$ in the range of $F_{D}$ that we have investigated,
and at $C = -1$ all the disks move in unison.
The fact that these phases can occur
over a range of relative drives indicates that 
such phases may be general to
other systems in which the particles are not exactly oppositely driven
but have differences in their relative driving.
These differences could arise for particles
with different drag coefficients, different coupling to a substrate,
different amounts of charge,
different shapes, and so forth.
Such effects could also be realized for
active matter systems
where one species couples to an external
drift force or where only one species is active.
There are already examples of mixtures
of active and non-active particles that undergo
phase separation \cite{39}.  
Another interesting feature of this system at $C = 0$ is that certain
regimes such as phase II flow exhibit a time dependent resistance.
When the drive is initially applied, the sample enters a high resistance
state in which many collisions occur,
but over time the disks approach a low resistance state
as the number of collisions are reduced.
This suggests that if finite temperature were included, then
when the drive is shut off there would be a finite time during which
the memory of the low-resistance organized state is 
preserved.
Such a system would have features similar to those of a memristor \cite{40}.     

\section{Summary}
We have numerically investigated a two-dimensional disk system in which two
disk species are driven in opposite directions.
We characterize the system by measuring the
average velocity of one disk species as a function of drive
to create a velocity-force curve analogous to what is observed in
driven systems with quenched disorder. 
The density $\phi$ is the 
total area coverage of the disks, and for $\phi > 0.55$ we
identify four dynamical phases: 
a crystalline jammed phase I, 
a fully phase separated state II
where the particles are not in contact but exhibit six-fold ordering, a 
strongly fluctuating liquid phase III where continuous collisions are
occurring, and a laning phase IV or smectic
state where collisions are absent.
The transitions between these
different phases are associated with
jumps or dips in the velocity-force curves and the differential mobility 
along with global changes in the disk configurations. 
At the transition into phase III we find
negative differential mobility.
The transitions
also correlate with jumps in the amount of six-fold ordering.    
For $\phi < 0.55$, the system always organizes
into the laning phase IV and the velocity-force curves are linear.   
We vary the relative driving forces on the two species, and consider the case
where one species is driven while the other species
does not couple to the drive 
as well as the case where both species are driven in the same
direction with different relative forces.
We find that the same four driven phases found for the oppositely driven case
still occur.
By measuring the instantaneous disk
velocities,
we find that in phases II and IV, the disks organize into a freely flowing state
in which disk-disk collisions no longer occur.
We discuss how the transitions into phases II and IV may be related to the
irreversible-reversible transitions recently observed
for periodically sheared colloidal or granular systems, which
organize to a state where the collisions between particles are lost.   

\section{Acknowledgements}
This work was carried out under the auspices of the 
NNSA of the 
U.S. DoE
at 
LANL
under Contract No.
DE-AC52-06NA25396.



\footnotesize{

}
\end{document}